\documentclass{cimento}

\usepackage{graphicx}  % got figures? uncomment this

\usepackage[switch]{lineno}
\usepackage{lipsum}
\title{The Fermi Gamma Ray Sky: summary of recent Observations}
\author{G.~Principe \from{ins:x}\from{ins:y}\from{ins:z}, \textit{on behalf of the Fermi-LAT Collaboration}}
\instlist{\inst{ins:x}Dipartimento di Fisica, Universit\'a di Trieste, I-34127 Trieste, Italy
\inst{ins:y} Istituto Nazionale di Fisica Nucleare, Sezione di Trieste, I-34127 Trieste, Italy
\inst{ins:z} INAF - Istituto di Radioastronomia, I-40129 Bologna, Italy}
\begin{document}

\maketitle

\begin{abstract}
The \textit{Fermi} Gamma-ray Space Telescope was launched more than 13 years ago and since then it has dramatically changed our knowledge of the gamma-ray sky. 
With more than three billions photons from the whole sky, collected in the energy range between 20 MeV and more than 300 GeV, and beyond 6,000 detected sources, LAT observations have been crucial to improving our understanding of particle acceleration and gamma-ray production in astrophysical sources. In this review, I will review recent science highlights from the LAT. 
I will focus on the recent source catalog release, as well as on the main transient phenomena seen with the LAT with multi-wavelength and multi-messenger connection. 
\end{abstract}

\section{Introduction}
The Fermi Large Area Telescope (LAT) has been surveying the high-energy gamma-ray sky since 2008 \cite{2009ApJ...697.1071A}.

\subsection{Fermi-LAT structure and performances}
The LAT detects photons by conversion into electron-positron pairs. It observes more than 20\% of the sky at any instant, covering the entire sky every 3 hours, with an operational energy range from 20\,MeV to more than 300\,GeV. It is composed of three main instruments: a high-resolution converter tracker (for direction measurement of the incident gamma rays), a CsI(Tl) crystal calorimeter (for energy measurement) and an anti-coincidence detector to distinguish the background of charged particles \cite{2009ApJ...697.1071A}.

After more than 13.5 years from the beginning of the mission, the LAT is in good operating condition with no performance concerns \cite{2021ApJS..256...12A}. 
In particular, it recently reached the following remarkable achievements:
\begin{itemize}
    \item more than 75500 orbits completed since launch,
    \item more than 5000 days of science mission (since 2008 Aug. 4),
    \item LAT has 98.7\% uptime for Science mission,
    \item more than 1.5 billion LAT events publicly available at the FSSC (reached on Feb 28) with an average of about 4 photons/second (including Earth limb).
\end{itemize}

\section{Fermi-LAT source catalogs}
\begin{figure}
\centering
\includegraphics[width=13cm]{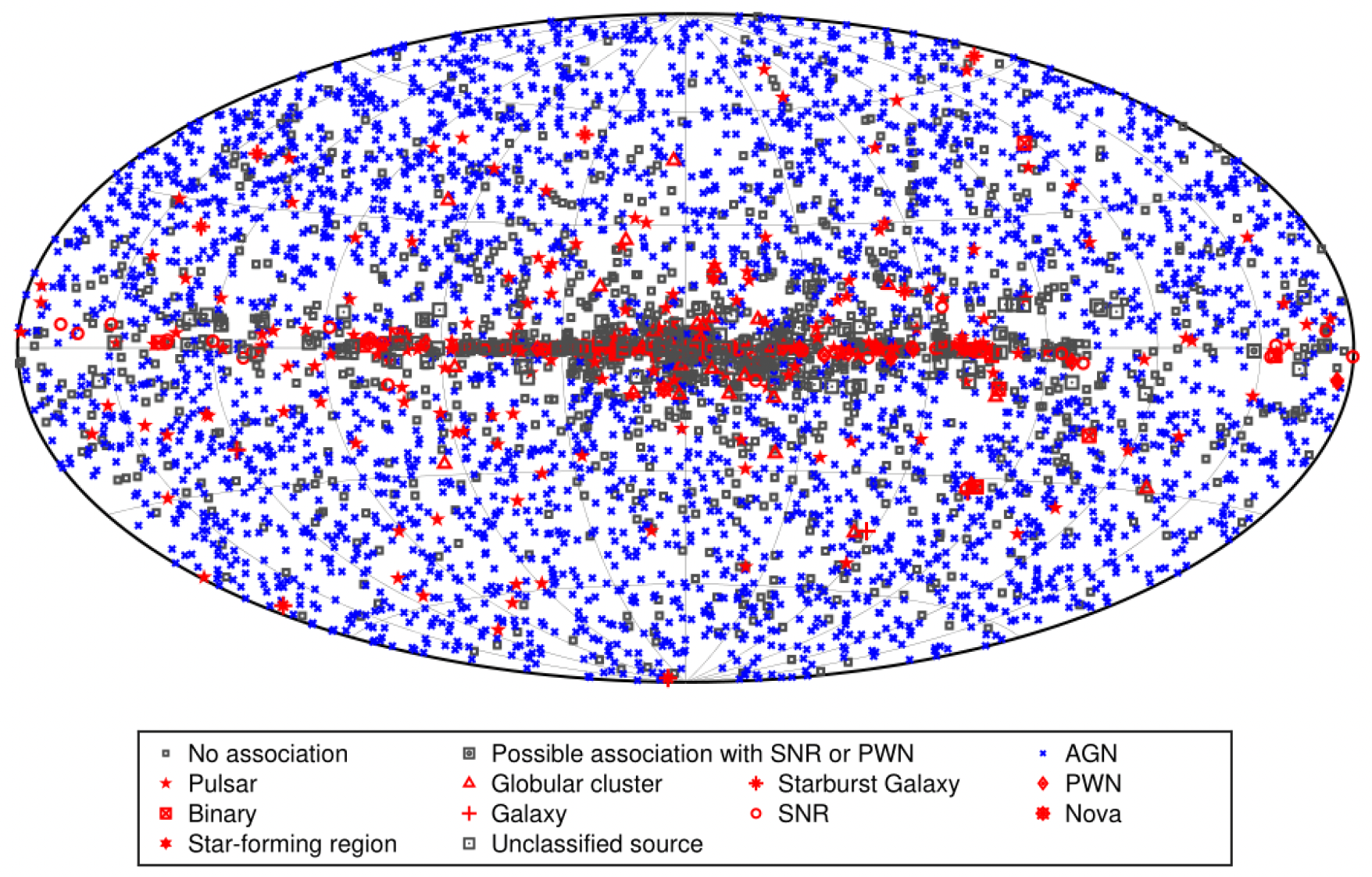}
\caption{\small \label{fig:4FGL_sky_map}
Sky map, in Galactic coordinates and Mollweide projection, showing the sources contained in the 4FGL catalog. All AGN classes are plotted with the same blue symbol for simplicity. Other associations to a well-defined class are plotted in red. Unassociated sources and sources associated to counterparts of unknown nature are plotted in black. The plot is taken from \cite{2020ApJS..247...33A}.}
\end{figure}

The list of steady sources detected by the LAT is an important
product of the \textit{Fermi} Collaboration activity.
The LAT Collaboration has published a succession of general source catalogs based on comprehensive analyses of LAT data between $\sim$100 MeV and 300 GeV energy range. This range has been extended in the latest catalogs to 50 MeV - 1 TeV. The fourth source catalog (4FGL) was obtained from the analysis of the first 8 years of LAT science data and contained 5064 sources \cite{2020ApJS..247...33A}. More than 3130 sources are classified as active galaxies of the blazar class, while 239 are pulsars.
Fig. \ref{fig:4FGL_sky_map} shows location and type of the sources contained in the 4FGL catalog.

An incremental version (4FGL-DR3, for Data Release 3) of the fourth Fermi-LAT catalog, containing more than 6500 gamma-ray sources, has been recently released \cite{2022arXiv220111184F}. It is based on the first twelve years of science data and makes use of an improved analysis with more sources presenting curved spectra and a more robust spectral parametrization for pulsars.

In addition to the general catalogs, which cover almost the entire LAT energy range, other catalogs have been built especially to investigate the lowest and highest LAT energies. The 1FLE catalog \cite{2018A&A...618A..22P} is based on 8.7 years of \textit{Fermi}-LAT data between 30 MeV and 100 MeV and contains almost 200 sources. 
Among the associated sources, 148 are extragalactic sources with the peculiarity that most BL Lacs in 1FLE are of the low-synchrotron peaked blazar type, which have softer spectra and higher redshifts with respect to those contained in the other LAT catalogs.
At high LAT energies, the 3FHL catalog \cite{2017ApJS..232...18A} covered the 10 GeV – 2 TeV energy range during the first 7 years of the \textit{Fermi} mission, with 1556 objects. The vast majority of detected sources (79\%) are associated with extragalactic counterparts at other wavelengths, including 16 sources located at very high redshift (z$>$2). 
Similarly, the LAT collaboration released some catalogs dedicated to specific source types, as the case of the first LAT Supernova Remnant Catalog \cite{2016ApJS..224....8A} or the forth catalog of Active Galactic Nuclei (4LAC) 
\cite{2020ApJ...892..105A}.

\section{Transient events seen by \textit{Fermi}}
Ever always increasing attention is given to the gamma-ray transient sources. 
Several are the types of transient sources detected by \textit{Fermi}-LAT, starting from pulsars with millisecond variability to gamma-ray binaries with year-long periodicity.
In order to investigate variability in this broad time range spanning more than ten decades ($10^{-2} - 10^{9}$ s) different pipelines have been produced.
In particular the latest pipeline release is the \textit{Fermi}-LAT light curve repository\footnote{https://fermi.gsfc.nasa.gov/ssc/data/access/lat/LightCurveRepository/about.html
}, which provides 3 day, 1 week and 1 month light curves for many 4FGL sources.
Similarly  a monthly transient catalog (1FLT) has recently been created, reporting the sources detected on monthly time intervals during the first decade of \textit{Fermi}-LAT operations \cite{2021ApJS..256...13B}.
In this section I will present some of the most recent major LAT results connected to gamma-ray transient events ranging from long to short time scales.

\subsection{AGN variability}
Two of the main open questions in AGN science are the jet formation and the origin of gamma-ray emission. 
A preferred attempt to investigate their origin is through coordinated MWL campaigns which are capable of resolving the fine structure of the source.
This can be applied in more evolved galaxies resolving the structure down to the core size (or even to the event horizon e.g.\cite{2019ApJ...875L...1E}) or on galaxies with recently-formed jets (i.e. $\sim$ 100-1000 years) which enables the investigation of the first moments of the jet formation \cite{2020A&A...635A.185P,2021MNRAS.507.4564P}. 
Another topic, related to the study of AGN variability, is the search for the origin of astrophysical neutrinos \cite{2018Sci...361.1378I}.

\subsubsection{Jet formation and the origin of gamma-rays}
In order to investigate these enigmas a broad multi-wavelength campaign was performed on M87 during April 2017 \cite{2021ApJ...911L..11E,2021RNAAS...5..221P}. This provided the most extensive, quasi-simultaneous, broadband spectrum of M87 yet taken (more than 17 decades in frequency), together with the highest ever resolution mm-VLBI images using the EHT from its 2017 April campaign. 
During the observations, the core of M87 appeared to be in a relatively low state, but clearly still dominating over the nearest knot HST-1. This provides the ideal observing conditions for a multiwavelength campaign combining data over a large range of spatial resolution. 
M87’s complex, broadband spectral energy distribution cannot be modeled by a single zone. 
Furthermore, while it was not possible to clearly understand where the VHE gamma-rays originate, it was possible to robustly rule out that they coincide with the EHT region for leptonic processes.

\subsubsection{Neutrino gamma-ray connection}
The association of a neutrino with a flaring blazar TXS\,0506+056 \cite{2018Sci...361.1378I} sparked interest in identifying additional counterparts.
So far, no other gamma-ray counterpart has been unambiguously identified \cite{2022icrc.confE.956G}.
However one source raised particular interest as a promising neutrino emitter: PKS\,1502+106; a bright FSRQ located at redshift of z$ = $1.84 \cite{2021ApJ...912...54R}.

\subsection{Solar flares}
Recently, the LAT collaboration released the first catalog of solar flares detected by \textit{Fermi}-LAT \cite{2021ApJS..252...13A}.
It contains 45 solar flares observed at 
energies between 30 MeV and 10 GeV over the years 2010-2018 (Solar cycle 24), increasing the total sample of $>$30 MeV detected solar flares by almost a factor of 10. 
In particular it reported the first detection of GeV emission from solar flares originating from active regions located behind the visible limb of the Sun.
While solar flares are dominated by electron emission, the LAT detected flares can be described by a curved model, either power-law with exponential cut off or a pion decay model, indicating the acceleration of protons and ions.

\subsection{Magnetar giant flares and Fast Radio Bursts}
\begin{figure}
\centering
\includegraphics[width=9cm]{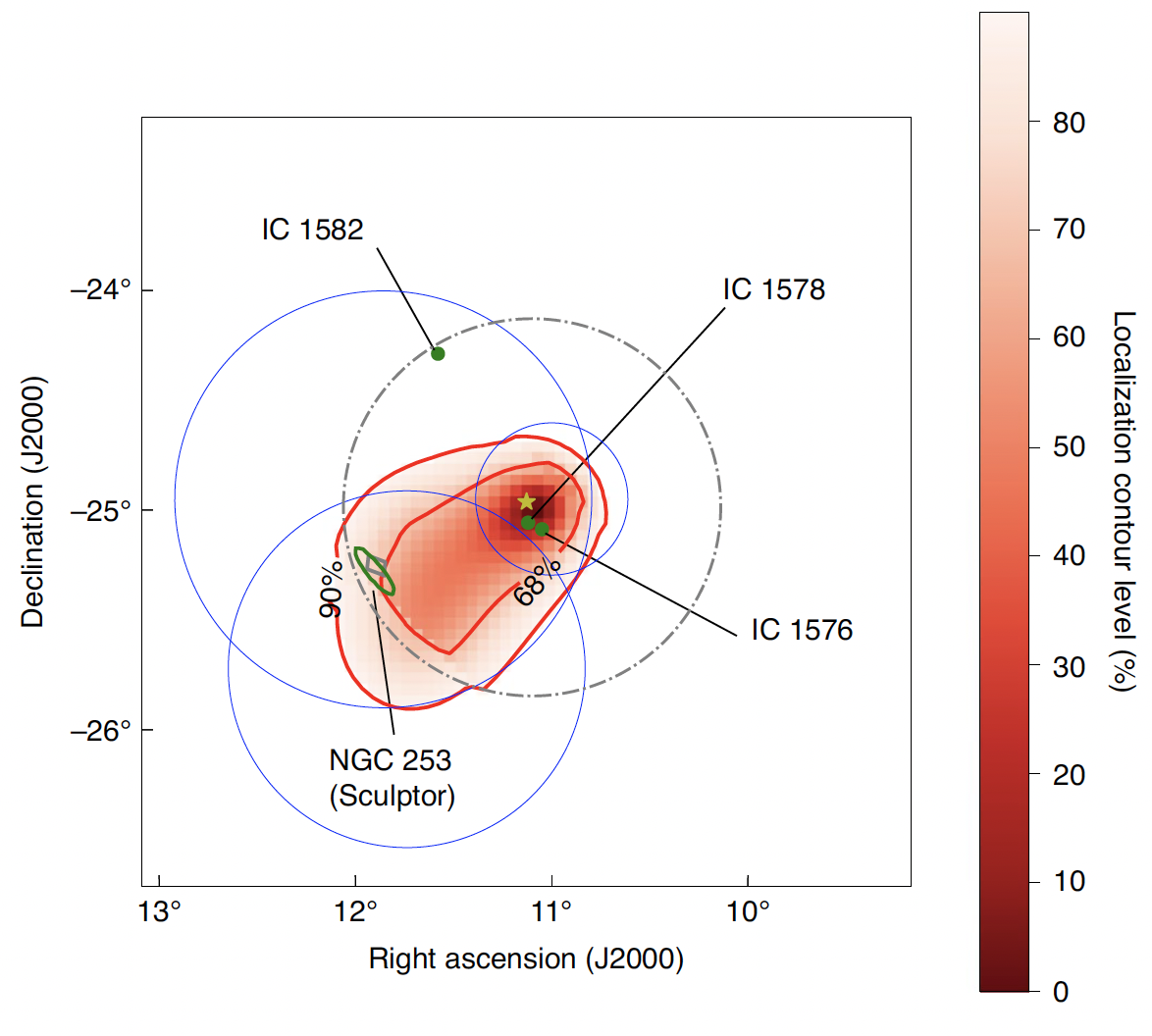}
\caption{\small \label{fig:sculptor}
Map of the localization contour probability for the photons associated to the Sculptor galaxy. The contours
encompassing a probability of 68\% and 90\% are displayed in red, while the
yellow star marks the location of the TS maximum. Galaxies from the NGC
2000 catalogue are shown as green disks, except for NGC 253 (Sculptor
galaxy), which is shown as an extended source. The grey box indicates the
localization provided by the IPN3. The blue circles indicate the 68\% containment of the point spread function
(PSF) for the three gamma rays probably associated with the flare.
 The plot is taken from \cite{2021NatAs...5..385F}.}
\end{figure}

Magnetars are strongly magnetized neutron stars with magnetic fields of $10^{13-15}$ G and periods of 0.1-10 s \cite{2015RPPh...78k6901T,2017ARA&A..55..261K}.
They can show rare outbursts (flare and pulsating tail) in X-rays and soft gamma-rays with luminosities around 10$^{44-47}$ erg s$^{-1}$.
These flares are likely caused by crustquakes induced by high magnetic fields.
On April 15th 2020 at 08:48:05.56 UTC, \textit{Fermi}-GBM was triggered by a bright burst \cite{2021Natur.589..207R}, which most likely originated in star-forming Sculptor Galaxy $D_L \sim$3.5 Mpc \cite{2021Natur.589..211S}.
\textit{Fermi}-LAT detected 3 photons (TS=29) from a source positionally consistent with the Sculptur galaxy, with a confidence level for the localization of 72\% \cite{2021NatAs...5..385F}. 
Fig. \ref{fig:sculptor} shows the map of the localization contour probability for the photons associated to the Sculptor galaxy. 

\noindent The probability of chance coincidence for this event has been estimated to be $<$2.9$\times $10$^{-3}$. In particular, a long delay of the first photon detected ($T_0=$20 s), compared
to typical values of $<$1 s between the GBM trigger time and
the LAT detection \cite{2019ApJ...878...52A}, is quite atypical for sGRB potentially indicating the importance of circumstellar material in VHE emission. The maximum energy observed for the photons detected from this source is 1.7 GeV.

Fast radio bursts (FRBs): are bright (Jy) and short-duration (few ms) radio pulses. Discovered just over a decade ago \cite{2007Sci...318..777L}, FRBs  are one of the newest astrophysical enigmas \cite{2019A&ARv..27....4P}. 
In April 2020, a first observation of a FRB from a Galactic magnetar giant flare (MGF), the soft gamma repeater SGR\,1935+2154, was firmly established established \cite{2020Natur.587...54C}. A coincident X-ray burst has been recorded by INTEGRAL and AGILE (\cite{2020ApJ...898L..29M,2021NatAs...5..401T}, respectively). Remarkably, this event showed for the first time that a magnetar can produce x-ray bursts in coincidence with FRB-like events. Motivated by the detection of GeV emission from a magnetar flare (Sculptor galaxy), \cite{2022icrc.confE.624P} in a ongoing project are performing the largest and deepest systematic search for gamma-ray emission from all the reported repeating and non-repeating FRBs using 12 years of \textit{Fermi}-LAT data.

\subsection{Gamma-ray bursts}
The \textit{Fermi}-LAT collaboration has recently released the second catalog of LAT-detected GRBs, covering the first 10 years of operations, from 2008 August 4 to 2018 August 4 \cite{2019ApJ...878...52A}. The catalog contains 186 GRBs; of these, 91 show emission in the range 30$-$100 MeV (17 of which are seen only in this band) and 169 are detected above 100 MeV. 
Among the detected bursts: 17 have been classified as short GRBs, and
169 as long GRBs.
The catalog presents the results for all 186 GRBs, in particular it contains information on onset, duration and temporal properties of each GRB, as well as spectral characteristics in the 0.1$-$100 GeV energy range.

\subsection{\textit{Fermi} Pulsar Timing Array}
Gravitational waves at very low frequencies (3-100 nHz) from merging of supermassive black hole (SMBH) binaries, may be detected through pulsar timing \cite{2015RPPh...78l4901L}.
In particular, they can be detected by monitoring the times of arrival of the steady pulses from each pulsar, which arrive earlier or later than expected due to the space-time perturbations.

\begin{figure}
\centering
\includegraphics[width=12cm]{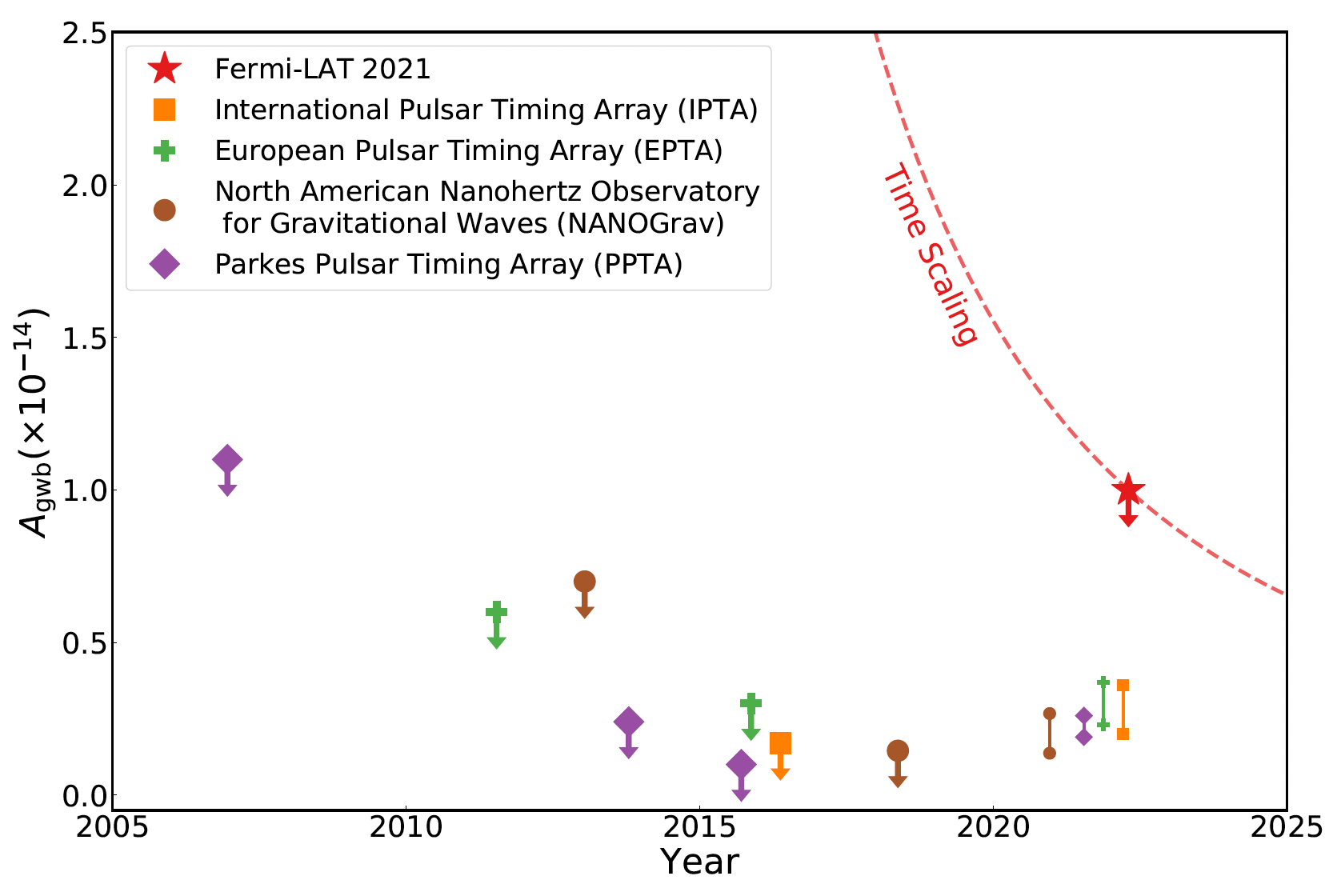}
\caption{\small \label{fig:fermi_pta}
Constraints on the gravitational wave background from radio and gamma-ray
PTAs. The \textit{Fermi}-LAT
95\% upper limit, 1.0$\times$10$^{-14}$, uses data obtained through January, 2021 and is shown as a red
star at an approximate publication date April, 2022. The dashed red line indicates the expected
scaling as the limit as a function of time. The plot is taken from \cite{2022Sci...376..521F}.}
\end{figure}

Using 12.5 years of LAT data to form a gamma-ray pulsar timing array (PTA) formed by
35 bright gamma-ray pulsars, it was possible to constrain the emission from the gravitational wave background (GWB) \cite{2022Sci...376..521F}. This first result obtained using gamma-ray data places a 95\% credible limit on the GWB characteristic strain of 1.0$\times$10$^{-14}$ yr$^{-1}$. 
Fig. \ref{fig:fermi_pta} shows the constraints on the gravitational wave background obtained with the gamma-ray photons detected by \textit{Fermi}-LAT in comparison with radio PTAs. 

\noindent The gamma-ray measurements are independent and have fewer sources of systematic uncertainty, providing a key check of radio observations and noise models.

\section{Conclusion}
Fermi LAT is working without major problems and continues to deliver exciting science results. After 13.5 years of data taking, discovery of (new) transient phenomena are particularly exciting. In particular, 
2020 marked the year of the detection of the first magnetar giant flare at GeV energies, while 2022 sees the publication of the results of the first GWB search with the \textit{Fermi}-LAT gamma-ray pulsar timing array.
The ground-breaking multi-messenger and multi-wavelenghts results summarised in this proceeding, highlight the important science still to come from \textit{Fermi}-LAT.
In addition, new observatories such as Square Kilometer Array\footnote{https://www.skatelescope.org} the Rubin Observatory \footnote{https://project.lsst.org/} and the Cherenkov Telescope Array\footnote{https://www.cta-observatory.org/} \cite{2019scta.book.....C} are either nearing completion or in active development and may overlap with the \textit{Fermi} mission. \textit{Fermi}-LAT is expected to centrally participate in coming major advances in multi-wavelength studies.

\acknowledgments
% Acknowledgement
\subsection*{ACKNOWLEDGMENTS}
The Fermi LAT Collaboration acknowledges generous ongoing support from a number of agencies and institutes that have supported both the development and the operation of the LAT as well as scientific data analysis. These include the National Aeronautics and Space Administration and the Department of Energy in the United States, the Commissariat à l'Energie Atomique and the Centre National de la Recherche Scientifique / Institut National de Physique Nucléaire et de Physique des Particules in France, the Agenzia Spaziale Italiana and the Istituto Nazionale di Fisica Nucleare in Italy, the Ministry of Education, Culture, Sports, Science and Technology (MEXT), High Energy Accelerator Research Organization (KEK) and Japan Aerospace Exploration Agency (JAXA) in Japan, and the K. A. Wallenberg Foundation, the Swedish Research Council and the Swedish National Space Board in Sweden.

Additional support for science analysis during the operations phase is gratefully acknowledged from the Istituto Nazionale di Astrofisica in Italy and the Centre National d'Etudes Spatiales in France. This work is performed in part under DOE Contract DE-AC02-76SF00515.

\end{document}